\documentclass{aa}

\usepackage[]{graphics}
\usepackage[]{rotating}
\usepackage[]{psfig}

\begin{document}

\thesaurus{03(11.04.4;11.04.1;11.01.1)}

\title{The central region of the Fornax cluster -- II. Spectroscopy and radial 
velocities of member and background galaxies}

\author {Michael Hilker \inst{1,2} \and Leopoldo Infante \inst{2} \and Gladys 
Vieira \inst{2} \and Markus Kissler-Patig \inst{3,1,4} \and Tom Richtler 
\inst{1}
} 

\offprints {M.~Hilker}
\mail{mhilker@astro.uni-bonn.de}

\institute{
Sternwarte der Universit\"at Bonn, Auf dem H\"ugel 71, 53121 Bonn, Germany
\and
Departamento de Astronom\'\i a y Astrof\'\i sica, P.~Universidad Cat\'olica,
Casilla 104, Santiago 22, Chile
\and 
UCO/Lick Observatory, University of California, Santa Cruz, CA 95064, USA
\and
Feodor Lynen Fellow of the Alexander von Humboldt Foundation
}

\date {Received --- / Accepted ---}

\titlerunning{The central region of the Fornax cluster -- spectroscopy of galaxies}
\authorrunning{M.~Hilker et al.}
\maketitle

\begin{abstract}

Radial velocities of 94 galaxies brighter than about $V_{\rm tot} = 20$ mag 
in the direction of the central Fornax cluster have been measured\footnote{
Table 2 containing the position, magnitude and velocity of all galaxies is
also available in electronic form at the CDS via anonymous
ftp to cdsar.u-strasbg.fr (130.79.128.5) or via 
http://cdsweb.u-strasbg.fr/Abstract.html}. 
Except for 8 Fornax members, all galaxies lie in
the background. Among the 8 members, there are 5 nucleated dwarf ellipticals
that are already listed in the FCC (Ferguson \cite{ferg89}). Two of the 
3 ``new''
members are very compact and have surface brightnesses comparable to globular 
clusters, however their luminosities are in the range of dwarf elliptical 
nuclei.

The measured line indices (especially Mg2, H$\beta$, and iron) of the brighter
of the compact objects suggest a solar metallicity, whereas
the fainter compact object as well as the dE,Ns have line
indices that are similar to those of old metal-poor globular clusters (GCs).
However, with these data it is not possible to clearly classify the
compact objects either as very bright globular clusters, isolated nuclei of 
dE,Ns, or even compact ellipticals.

A background galaxy
cluster at $z = 0.11$ has been found just behind the center of the Fornax 
cluster. This explains the excess population of galaxies reported in
Paper~I. The brightest galaxy of the background cluster lies only $1\farcm1$ 
south of NGC~1399 and is comparable in absolute 
luminosity with the central Fornax galaxy itself.


\end{abstract}

%%%%%%%%%%%%%%%%%%%%%%%%%%%%%%%%%%%%%%%%%%%%%%%%%%%%%%%%%%%%%%%%%%%%%%%%%%%
%%%%%%%%%%%%%%%%%%%%%%%%%%%%%%%%%%%%%%%%%%%%%%%%%%%%%%%%%%%%%%%%%%%%%%%%%%%
%%%%%%%%%%%%%%%%%%%%%%%%%%%%%%%%%%%%%%%%%%%%%%%%%%%%%%%%%%%%%%%%%%%%%%%%%%%

\section{Introduction}

In the center of galaxy clusters the various types of dwarf galaxies
differ in properties, such as their spatial distribution 
and luminosity function. Early-type dwarf ellipticals, especially nucleated
ones, are the most clustered, whereas late-type dwarfs are
preferentially found at the outskirts of clusters (e.g. review by
Ferguson \& Binggeli \cite{ferg94}, and references therein).
The different environmental properties probably reflect the result of the 
dynamical procresses during the formation and evolution epoch of the clusters.

Among low mass galaxies, a striking but rare subgroup
are the compact ellipticals (cEs), or M32-type galaxies. 
It is under discussion, whether environmental effects such as tidal stripping
have shaped their present appearance (e.g. Faber \cite{fabe}) or whether 
these galaxies just represent the low-luminosity
end of the giant ellipticals (Nieto \& Prugniel \cite{niet}). E.g.,
M32 itself is a companion of the Andromeda galaxy (M31). The non-existence of
any globular cluster (GC) in this galaxy suggests that previously present
GCs might have been captured by M31 during past close passages.
Whether the interaction with M31 also has formed the compact shape of M32
remains unclear.

At the distance of the nearest galaxy clusters, the classification of compact 
objects such as M32 just by morphological properties is nearly impossible.
Their high central surface brightness and de Vaucouleurs surface brightness
profile make them indistinguishable from 
background ellipticals. In the Fornax Cluster Catalog (FCC)
of possible members and likely background galaxies, Ferguson (\cite{ferg89}) 
lists 131 candidate cEs.
Recently, Drinkwater et al.~(\cite{drin97}, see also Drinkwater \& Gregg 
\cite{drin98}) have measured
radial velocities of 67 of them. No one is member of the Fornax cluster.
The fact that very few cEs are found indicates that tidal stripping
does not play a major role in their formation in rich environments.
However, the member and background galaxy catalogs of Ferguson are limited in 
the angular diameter of the objects. Especially, galaxies with diameters 
comparable to M32 itself and smaller compact ellipticals are beyond the 
resolution limit of Ferguson's sample.

A second type of objects in the galaxy cluster populations that could easily 
be missed by morphological
classifications are nuclei of tidally disrupted nucleated dwarf 
ellipticals (dE,Ns). Numerical simulations by Bassino et al. (\cite{bass})
reveal that such nuclei can survive the dissolution in the gravitational field 
even during the entire
lifetime of the universe and would appear as luminous globular cluster-like
objects. The nuclear magnitudes of all Virgo dE,Ns (Binggeli \& Cameron 
\cite{bing91}),
for example, fall indeed in the magnitude -- surface brightness sequence defined
by the globular clusters (e.g. Binggeli \cite{bing94}).
At the distance of the Fornax cluster, these objects would hardly be 
resolved and can only be uncovered by spectroscopic observations. 

In the first paper (Hilker et al.~\cite{hilk}, hereafter Paper~I) a 
galaxy catalog with photometric properties
and surface brightness profiles for galaxies in selected fields of the central
Fornax cluster has been presented. An excess population of 
galaxies near NGC~1399, the central galaxy of the cluster, has been found as
compared to the other Fornax fields and to absolute background fields.
The photometric analysis
has shown that most of the excess galaxies have sizes and surface 
brightnesses which are more
typical for background spirals and ellipticals than for dwarf ellipticals. 
However, as discussed before, photometric properties alone 
are not sufficient to distinguish between background galaxies and high 
surface brightness dwarf galaxies in the Fornax cluster.
This paper presents redshift determinations of a bright sub-sample
of our photometric catalog (Paper~I) to investigate the nature of the 
mentioned excess galaxies. Furthermore, line indices for the 
objects that have been identified as Fornax members were measured.

In the following the expression `radial velocity' has been used for the
measurement of $cz$ instead of redshift, being aware of the fact that
the true radial velocity for high $z$ differs from $cz$ depending on the
applied cosmological model.

Previous radial velocity measurements of galaxies in the Fornax cluster
brighter than $B_T = 15.5$ mag
were presented by Jones \& Jones (\cite{jone}), Lauberts (\cite{laub}) and 
Richter \& Sadler (\cite{rich}).
They are compiled in the Fornax Cluster Catalog (FCC) by 
Ferguson (\cite{ferg89}).
Except for the giant galaxies, there are only two galaxies that overlap with
our sample: NGC~1396 and FCC~222, two bright dE,Ns.
Held \& Mould (\cite{held}) took spectra of 10 dE,Ns in
the Fornax cluster; one of these is in common with our sample.

Sect.~2, 3, and 4 give a detailed description of the observations,
data reduction, and velocity determination. The resulting radial velocities
and the analysis of individual objects are presented in Sect.~5. 
The main results are summarized in Sect.~6.
%
%%%%%%%%%%%%%%%%%%%%%%%%%%%%%%%%%%%%%%%%%%%%%%%%%%%%%%%%%%%%%%%%%%%%%%%%%%%
%%%%%%%%%%%%%%%%%%%%%%%%%%%%%%%%%%%%%%%%%%%%%%%%%%%%%%%%%%%%%%%%%%%%%%%%%%%
%%%%%%%%%%%%%%%%%%%%%%%%%%%%%%%%%%%%%%%%%%%%%%%%%%%%%%%%%%%%%%%%%%%%%%%%%%%
%
\section{Observations}

\begin{figure}
\psfig{figure=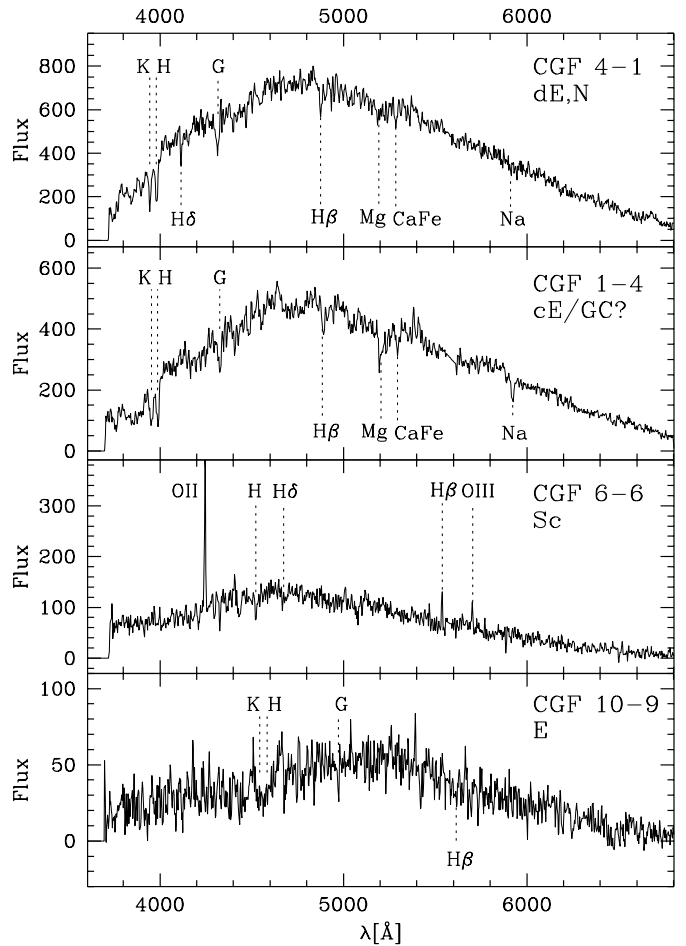,height=12.3cm,width=8.6cm
,bbllx=13mm,bblly=66mm,bburx=139mm,bbury=250mm}
\vspace{0.4cm}
\caption{Typical spectra after subtraction of the sky spectra. The uppermost
panel shows the nucleated dwarf elliptical NGC~1396. In the panel below the
spectrum of the compact nucleus-like Fornax member CGF~1-4 is shown.
In the third panel a typical emission line spectrum of a background spiral
is plotted. The spectrum in the bottom panel
has a signal-to-noise at the limit where velocity measurements are still 
feasible
}
\end{figure}

The observations have been performed with the 2.5m DuPont telescope at
the Las Campanas Observatory, Chile, during the nights of 7--9 December, 1996.
The multi-fiber spectrograph from Shectman (\cite{shec}) has been used. The 
field of view
is $1.5\times1.5$ degree. The aperture size of each fiber is 2 arcsec in
diameter on the sky. The fibers were connected with a Boller \& Chivens
spectrograph coupled to a 2DFrutti detector (2DF).
With a blaze angle of $9^\circ45\arcmin$ we got a
spectral range of 3800\AA - 6800\AA. The spectra are projected to a
$1520\times1024$ pixel area, with a dispersion of $\simeq$ 2--3 \AA$\cdot$
pixel$^{-1}$ and a final resolution of $\simeq$5\AA.

From our catalog 125 galaxies brighter than $V = 20.0$ mag were selected.
Accurate positions have been obtained by using reference stars
from the Guide Star Catalog. Astrometric
solutions yielded positions with accuracies better than $0\farcs3$.
In addition, three bright dE,Ns have been included from 
the FCC (Ferguson \cite{ferg89}), namely FCC 188, FCC 222, and
FCC 274. Their positions are not as accurate as those of the CCD sample. Small
positional deviations for these objects resulted in quite low signal-to-noise 
spectra despite the high central surface brightnesses of their nuclei.
The whole sample was divided in two sets of positions in order to avoid
overlaps in the fiber configuration. In spite of this, 13 positions could not
be taken. About 20 to 30 fibers per set were positioned as sky fibers to random 
positions in the field.
The integration times of the first and second set were 4 hours
and 4.5 hours respectively. For the wavelength calibration comparison
HeNe spectra have been obtained at the two positions before and after
the long integrations.
Additionally, long flatfield exposures have been taken. We have chosen
different grating angles between $8\degr45\arcmin$ and $10\degr45\arcmin$ to
get an homogenous illumination of the whole spectral range.
%
%%%%%%%%%%%%%%%%%%%%%%%%%%%%%%%%%%%%%%%%%%%%%%%%%%%%%%%%%%%%%%%%%%%%%%%%%%%
%%%%%%%%%%%%%%%%%%%%%%%%%%%%%%%%%%%%%%%%%%%%%%%%%%%%%%%%%%%%%%%%%%%%%%%%%%%
%%%%%%%%%%%%%%%%%%%%%%%%%%%%%%%%%%%%%%%%%%%%%%%%%%%%%%%%%%%%%%%%%%%%%%%%%%%

\section{Reductions}

All reduction steps have been performed with packages under IRAF.
We followed essentially the method described by Way et al.~(\cite{way}) with
the following basic reductions steps.

All spectra are arranged side by side on the 2DF x-y-pixel image, the y 
direction being the direction of dispersion.
Since the spectra have a strong curvature on the
x-y-pixel plane, they have to be straightened first. We fitted a 
transformation of the plane in x direction by tracing the features in the
combined image of the flatfield exposures. Similarily, the features in the
combined calibration lamp exposure were fitted for the y transformation.
All images were straightened by these two transformations.
(This procedure uses the IRAF tasks {\it identify, reidentify, fitcoords}
and {\it transform} in the {\it noao.twodspec.longslit} package.)

The extraction of the single spectra, tracing of the apertures, fitting
of the flatfield, wavelength calibration with the two arc spectra, and
subtraction of the sky spectra have been done with the {\it dohydra}
package. The residual sky lines in the spectra were cleaned with the
{\it lineclean} task. The typical rms error of the solution for the wavelength 
calibration is less than 0.3\AA.

Figure 1 shows 4 typical spectra of different signal-to-noise, with some
reference lines indicated.

\section{Radial velocity measurements}

The radial velocities $cz$ have been determined by two different, independent 
methods:
first, ``automatic'' identification of the absorption and/or emission
lines and fourier cross correlation with template spectra using the {\it rvsao}
package; second, ``by eye'' identification of absorption lines and
computation of $cz$ with the task {\it rvidlines}.
Both methods were applied to all measured spectra.

\begin{figure}
\psfig{figure=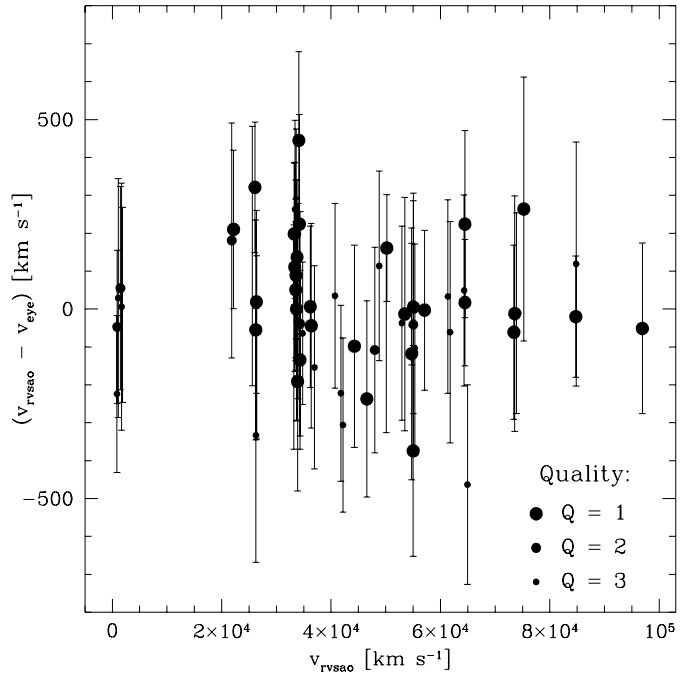,height=8.6cm,width=8.6cm
,bbllx=9mm,bblly=65mm,bburx=195mm,bbury=246mm}
\vspace{0.4cm}
\caption{This plot shows the differences of the two independent methods
of velocity determinations. The rms error of the scatter of the most
reliable identifications (Q = 1 and 2) is about 150 km~s$^{-1}$
}
\end{figure}

A number of our spectra show O\,{\sc ii}, H$\beta$, and O\,{\sc iii} in 
emission. Emission lines were removed form the spectra and a cross-correlation
analysis was carried out. Radial velocities were also independently measured
for these emission lines.

For the first method high signal-to-noise spectra of selected galaxies 
(NGC~1407, NGC~1426, and NGC~1700) were employed as templates in the
the cross-correlation analysis. These templates were observed with the
same instrument. A synthetic spectrum, as in  Way et al.~(\cite{way}), was also used.
All templates were rebinned to the same resolution as the fiber spectra and
were corrected for heliocentric velocity. 
The task {\it xcsao} crosscorrelates the emission-line-cleaned object spectra
with the template spectra and generates a reliability factor called
the R value which depends on the amplitude of the cross correlation peak. 
A low R value (R$<4$) indicates that one should control the 
results by eye. The ``best'' velocity, as judged by the highest R value
was adopted.
In cases where strong emission lines, especially the O\,{\sc ii} line, 
were present we calculated in addition their velocities with the task 
({\it emsao}).
In some cases where the absorption lines were very weak, but
the emission lines clearly visible, the estimated emission line velocity
could be used to improve the absorption line detection with {\it xcsao}. 
The absorption line velocities $v_{\rm cross}$ as well as the emission line
velocities $v_{\rm em}$ are presented in Table 2. 

\begin{table}
\caption{Objects whose spectra have a dominant contribution
by light of a close Fornax giant galaxy}
\begin{flushleft}
\begin{tabular}{llcrclc}
\hline\noalign{\smallskip}
Object id. & giant  & light & $v_{\rm giant}$ & Q & $v_{\rm obj}$ & Q \\
 & NGC & ratio & [km~s$^{-1}$] & & [km~s$^{-1}$] & \\
\noalign{\smallskip}
\hline\noalign{\smallskip}
CGF~9-9 & 1374 & 0.79 & 1414$\pm$71 & 2 & 38490$\pm$74 & 3 \\
CGF~8-3 & 1379 & 7.33 & 1504$\pm$49 & 1 & & \\
CGF~6-5 & 1387 & 1.65 & 1277$\pm$38 & 1 & & \\
CGF~3-11    & 1399 & 1.62 & 1541$\pm$57 & 3 & 34225$\pm$70 & 1 \\
CGF~4-5    & 1399 & 4.00 & 1443$\pm$45 & 1 & & \\
CGF~4-8    & 1399 & 0.26 & 1281$\pm$97 & 2 & 33531$\pm$52 & 2 \\
CGF~1-13    & 1404 & 0.02 & 2253$\pm$85 & 4 & 44267$\pm$105 & 4 \\
CGF~10-1  & 1427 & 1.01 & 1349$\pm$40 & 1 & 16409$\pm$40 & 2 \\
CGF~10-14 & 1427 & 1.28 & 1483$\pm$48 & 1 & & \\
\noalign{\smallskip}
\hline
\end{tabular}
\end{flushleft}
\end{table}

The second method is based on the following steps:
the Balmer jump and the close K and/or H lines were identified by eye and
marked for the IRAF task {\it rvidlines} which compares the marked lines
with a list of absorption lines. All identified lines with a sufficient
signal-to-noise were used for the velocity calculation. The centering
algorithm used for the position of the lines calculates the wavelength where
the total flux of the absorption feature is divided into equal halfs.
The resulting velocities $v_{\rm rvid}$ are presented in Table 2.

The results of both methods were compared. In 18 of the 115 spectra the
signal-to-noise was too low to determine a velocity by either of the methods.
In 68 spectra the velocities of
both methods agree within $\pm$500 km~s$^{-1}$. Figure 2 shows the
differences of
the two determinations. The standard deviation 
is about 130 km~s$^{-1}$, comparable to the rms errors of the fit to individual
velocity determinations.
One of our galaxies, the nucleated dwarf elliptical NGC~1396, was measured
in both fiber sets. The signal-to-noise is one of the highest in our sample.
The velocities of the ``automatic'' identification
agree very well: 865$\pm$17 and 871$\pm$32 km~s$^{-1}$. The ``by eye'' 
identification
yields 870$\pm$44 and 918$\pm$52 km~s$^{-1}$. Within the errors all values are
consistent and agree with the published value by Jones \& Jones (\cite{jone}):
894$\pm$120 km~s$^{-1}$.

In 9 cases where the object is
located very close to a Fornax elliptical or S0 galaxy, most probably
the light, and thus the velocity of the giant was measured instead that of 
the object.
In 5 of these spectra a second velocity was found indeed. In Table 1 we give 
the name of the Fornax galaxy, the ratio of the Fornax giant galaxy to target 
object 
light in a 2 arcsec aperture, the velocity of both components,
and the identification quality as defined in the next section.

\section{Results}

\begin{table*}
\caption{Radial velocities of galaxies in and behind the Fornax cluster}
\begin{flushleft}
\begin{tabular}{llrrrrrrrrrrc}
\noalign{\smallskip}
\hline
Object Id. & Type & RA(2000) & Dec(2000) & V & $v_{\rm cross}$ & R & $v_{\rm rvid}$ &
$v_{\rm em}$ & $v_{\rm helio}$ & $error_v$ & Q \\
\noalign{\smallskip}
\hline
\noalign{\smallskip}
CGF~9-5  & S/BCD?   & 3:34:57.23 &-35:12:24.7 & 19.06 &       &      & & 16833 & 16833 &  95 & 3\\
CGF~9-2  & S        & 3:34:59.79 &-35:09:58.1 & 18.24 &       &      & 18649 & 18058 & 18354 &  77 & 4\\
CGF~9-3a & Sc/I?    & 3:35:02.47 &-35:11:49.6 & 18.39 & 10939 &  4.5 & 10396 &      & 10649 &  63 & 3\\
CGF~9-3b & Sc/I?    & 3:35:02.47 &-35:11:49.6 & 18.39 &       &      & & 21205 & 21205 &  75 & 3\\
CGF~9-7  & E        & 3:35:06.52 &-35:10:54.1 & 19.28 & 61733 &  5.0 & 61794 &      & 61763 &  61 & 1\\
CGF~9-9  & S        & 3:35:14.57 &-35:14:04.6 & 19.94 &       &      & 38490 &      & 38490 &  74 & 3\\
CGF~9-1  & S0       & 3:35:25.07 &-35:16:09.6 & 17.74 & 50075 & 11.7 & 50154 &      & 50115 &  45 & 1\\
CGF~8-2  & E/S0?    & 3:36:13.38 &-35:26:04.9 & 19.41 &       &      & 21077 &      & 21077 &  80 & 3\\
CGF~8-4  & Sc       & 3:36:27.47 &-35:26:02.3 & 19.95 &       &      & & 94531 & 94531 &  40 & 2\\
CGF~6-3  & Sbc      & 3:36:44.44 &-35:33:48.1 & 18.88 & 42022 &  3.7 & 42328 & 41732 & 42175 &  74 & 2\\
CGF~6-4  & S/Irr?   & 3:36:45.15 &-35:31:48.5 & 19.04 & 52034 &  3.1 & & 51420 & 51761 &  77 & 3\\
CGF~6-6  & Sc       & 3:36:48.98 &-35:31:52.7 & 19.10 &       &      & 41898 & 41676 & 41787 &  62 & 3\\
CGF~6-2  & Sc       & 3:36:56.93 &-35:28:21.4 & 18.51 & 50228 &  5.6 & 50105 & 50380 & 50186 &  44 & 1\\
CGF~6-8  & Sb(LSB?) & 3:37:00.58 &-35:32:12.5 & 19.97 & 18539 &  2.6 & 19523 &      & 19031 &  67 & 4\\
CGF~6-1  & Sc       & 3:37:01.46 &-35:33:58.9 & 18.21 & 34794 &  6.1 & 34826 & 34729 & 34794 &  58 & 3\\
FCC 188 & dE,N     & 3:37:04.44 &-35:35:19.3 & 16.10 &   945 &  5.1 & &      &   945 &  67 & 3\\
CGF~3-9  & S(B)a?   & 3:38:01.97 &-35:32:37.5 & 18.91 & 33656 & 10.1 & 33549 &      & 33594 &  47 & 1\\
CGF~4-3  & S        & 3:38:05.25 &-35:22:39.4 & 18.88 &       &      & 46324 &      & 46324 &  78 & 4\\
CGF~3-5  & E        & 3:38:05.98 &-35:32:19.1 & 18.05 & 33731 & 11.2 & 33760 &      & 33745 &  37 & 1\\
CGF~4-1  & dEN      & 3:38:06.54 &-35:26:24.4 & 14.88 &   868 &  8.0 & 918 &      &   882 &  39 & 1\\
CGF~3-7  & Sb       & 3:38:06.58 &-35:29:08.9 & 18.52 & 21933 &  6.4 & 21744 & 21726 & 21835 &  29 & 1\\
CGF~4-4  & E        & 3:38:09.43 &-35:25:43.0 & 18.92 & 40754 &  3.7 & 40696 & 40740 & 40714 &  60 & 2\\
CGF~4-6a & S        & 3:38:15.02 &-35:27:33.5 & 19.16 & 52835 &  2.8 & 52958 & 53006 & 52939 & 111 & 3\\
CGF~4-6b & S        & 3:38:15.02 &-35:27:33.5 & 19.16 &       &      & 28376 &      & 28376 &  94 & 3\\
CGF~3-13a & Sc(pec)  & 3:38:16.61 &-35:30:11.0 & 19.48 &       &      & 55208 & 54834 & 55021 &  98 & 3\\
CGF~3-13b & Sc(pec)  & 3:38:16.61 &-35:30:11.0 & 19.48 &       &      & 38944 &      & 38938 &  79 & 4\\
CGF~4-8  & Sb?      & 3:38:17.78 &-35:26:43.7 & 19.95 & 33569 &  4.4 & 33492 &      & 33531 &  52 & 2\\
CGF~3-2  & dEN      & 3:38:18.71 &-35:31:52.1 & 17.21 &  1625 &  4.4 & 1691 &      &  1694 &  85 & 1\\
CGF~3-11  & S?       & 3:38:21.30 &-35:28:50.3 & 19.05 &       &      & & 52788 & 52788 &  85 & 4\\
CGF~3-6  & Sa       & 3:38:21.79 &-35:32:58.0 & 18.33 & 61415 &  7.4 & 61348 &      & 61365 &  53 & 1\\
CGF~3-12  & Sc?      & 3:38:23.07 &-35:28:07.2 & 19.27 & 34219 &  4.3 & 34113 & 34455 & 34225 &  70 & 1\\
CGF~4-2  & Sa       & 3:38:23.49 &-35:24:06.4 & 17.46 & 33946 &  8.8 & 33953 &      & 33964 &  62 & 1\\
CGF~3-4  & ring?    & 3:38:25.83 &-35:33:28.4 & 17.95 & 54689 &  8.8 & 54797 & 54663 & 54738 &  57 & 1\\
CGF~1-11  & S0       & 3:38:28.89 &-35:28:24.6 & 18.74 & 33886 & 10.9 & 33716 &      & 33785 &  40 & 1\\
CGF~1-1  & E(cD)    & 3:38:29.16 &-35:28:08.6 & 16.28 & 33440 & 18.1 & 33299 &      & 33355 &  44 & 1\\
CGF~2-3  & E/S0     & 3:38:32.17 &-35:24:30.2 & 18.35 & 33718 & 18.9 & 33509 &      & 33608 &  46 & 1\\
CGF~1-15  & E        & 3:38:32.96 &-35:30:12.4 & 19.09 & 36893 &      & 37047 &      & 36970 &  72 & 2\\
CGF~1-16  & S0       & 3:38:35.16 &-35:27:40.7 & 19.11 & 33583 &  7.7 & 33280 &      & 33412 &  48 & 1\\
CGF~1-18  & Sb       & 3:38:35.93 &-35:31:13.5 & 19.16 &       &      & 48747 & 48861 & 48804 &  72 & 3\\
CGF~2-9  & Sbc      & 3:38:40.58 &-35:20:24.5 & 19.16 & 33804 &  4.4 & 33875 & 33702 & 33840 &  62 & 2\\
CGF~1-9  & S0?      & 3:38:40.93 &-35:27:26.7 & 18.48 & 34321 &  6.4 & 34360 & 34138 & 34341 &  65 & 1\\
CGF~1-12  & Sa?      & 3:38:41.61 &-35:27:25.1 & 18.83 & 34257 &  7.6 & 34391 & 34194 & 34324 &  49 & 1\\
CGF~1-3  & Sbc      & 3:38:44.75 &-35:27:01.3 & 17.48 & 33143 &  6.8 & 33224 & 33157 & 33187 &  55 & 1\\
CGF~1-14  & S0       & 3:38:45.47 &-35:29:53.9 & 19.08 & 33354 & 11.3 & 33156 &      & 33255 &  34 & 1\\
CGF~2-2  & E(pec)   & 3:38:45.97 &-35:22:52.3 & 17.82 & 33798 &  8.9 & 33964 & 33785 & 33869 &  58 & 1\\
CGF~2-18  & E        & 3:38:46.15 &-35:22:35.2 & 19.91 & 84708 &  4.3 & 84758 &      & 84748 &  55 & 3\\
CGF~2-17  & Irr(LSB) & 3:38:46.44 &-35:20:55.9 & 19.82 & 26504 &  3.3 & 25721 & 26422 & 26092 &  83 & 3\\
CGF~1-21  & E/S0     & 3:38:48.20 &-35:33:18.9 & 19.61 & 64630 &  7.4 & 64358 &      & 64470 &  56 & 1\\
CGF~1-13  & Sc       & 3:38:49.07 &-35:33:44.1 & 18.97 &       &      & 44316 & 44218 & 44267 & 105 & 3\\
CGF~2-6  & S0/E     & 3:38:49.10 &-35:25:30.3 & 18.62 & 33726 &  5.2 & 33675 & 33471 & 33675 &  65 & 1\\
CGF~1-22  & S/Irr?   & 3:38:49.21 &-35:29:19.6 & 19.74 & 82506 &  2.2 & & 82343 & 82435 & 168 & 4\\
CGF~1-17  & E        & 3:38:50.31 &-35:30:56.9 & 19.14 & 73578 &  9.3 & 73590 &      & 73584 &  56 & 1\\
CGF~2-15  & E(pec?)  & 3:38:51.19 &-35:22:15.8 & 19.55 & 73917 &  8.2 & 73894 & 74159 & 73889 &  54 & 1\\
CGF~2-14  & E/S0     & 3:38:51.50 &-35:24:22.2 & 19.48 & 34341 &  4.1 & 33874 &      & 34097 &  60 & 3\\
CGF~2-8  & Sb(pec?) & 3:38:51.76 &-35:22:18.0 & 18.97 & 84930 &  5.9 & 84749 & 84587 & 84809 &  85 & 2\\
CGF~2-16  & Irr      & 3:38:53.08 &-35:24:48.7 & 19.80 &       &      & & 33540 & 33540 &  88 & 4\\
CGF~1-4  & cE/GC?   & 3:38:54.05 &-35:33:33.9 & 17.87 &  1523 & 23.9 & 1457 &      &  1485 &  38 & 1\\
CGF~1-19  & E        & 3:38:54.48 &-35:27:55.9 & 19.55 & 73372 &  4.3 & 73461 &      & 73431 &  62 & 3\\
\noalign{\smallskip}
\hline
\end{tabular}
\end{flushleft}
\end{table*}

\begin{table*}
\begin{flushleft}
\begin{tabular}{llrrrrrrrrrc}
\multicolumn{12}{l}{\bf Table 2. (continued)}\\
\noalign{\smallskip}
\hline
Object Id. & Type & RA(2000) & Dec(2000) & V & $v_{\rm cross}$ & R & $v_{\rm rvid}$ &
$v_{\rm em}$ & $v_{\rm helio}$ & $error_v$ & Q \\
\noalign{\smallskip}
\hline
\noalign{\smallskip}
CGF~1-7  & S0/Sb?   & 3:38:54.70 &-35:31:04.6 & 18.36 & 53478 &  6.6 & 53470 & 53437 & 53464 &  58 & 1\\
CGF~1-6  & E        & 3:38:57.22 &-35:32:33.8 & 18.34 & 55066 & 12.2 & 55061 &      & 55064 &  48 & 1\\
CGF~1-10  & S0/E     & 3:38:59.58 &-35:30:31.7 & 18.69 & 64581 &  6.3 & 64460 &      & 64469 &  41 & 1\\
CGF~2-7  & Irr      & 3:39:00.98 &-35:20:29.2 & 18.73 & 33639 &  5.1 & 33517 & 33497 & 33542 &  58 & 1\\
CGF~2-4  & S0?      & 3:39:01.77 &-35:22:56.6 & 18.51 & 58255 &  2.8 & & 58030 & 58143 & 123 & 4\\
CGF~5-8  & S?       & 3:39:07.65 &-35:29:59.7 & 19.60 & 64703 &  4.1 & 65166 & 64403 & 64935 &  67 & 3\\
FCC 222 & dE,N     & 3:39:13.32 &-35:22:10.9 & 15.60 &   740 &  8.0 & 964 &      &   852 &  52 & 1\\
CGF~5-13  & Irr      & 3:39:18.44 &-35:24:00.5 & 20.03 &       &      & 75110 & 75374 & 75242 &  89 & 2\\
CGF~5-7  & Sa       & 3:39:18.61 &-35:27:42.9 & 19.41 & 64349 &  3.9 & 64300 &      & 64324 &  67 & 3\\
CGF~5-2  & E        & 3:39:22.44 &-35:25:30.9 & 18.06 &       &      & & 13904 & 13904 &  77 & 2\\
CGF~5-5  & S        & 3:39:22.63 &-35:25:29.1 & 19.36 & 26220 & 10.6 & 26261 & 26192 & 26233 &  57 & 1\\
CGF~5-10  & E        & 3:39:25.66 &-35:23:50.2 & 19.74 & 26355 &  4.6 & 26468 &      & 26418 &  69 & 1\\
CGF~5-1  & S0       & 3:39:26.01 &-35:24:14.5 & 17.31 & 26404 & 19.5 & 26339 &      & 26352 &  37 & 1\\
CGF~5-6  & E/S0     & 3:39:29.37 &-35:23:50.3 & 19.39 &       &      & 53106 &      & 53106 &  74 & 4\\
CGF~5-11  & BCD?     & 3:39:29.60 &-35:26:52.0 & 19.76 & 26342 &  2.8 & 25912 & 26125 & 26073 & 124 & 3\\
CGF~5-9  & Sc       & 3:39:33.67 &-35:27:06.8 & 19.67 &       &      & 26432 & 26099 & 26266 &  93 & 2\\
CGF~5-12  & S?       & 3:39:33.68 &-35:25:52.9 & 19.93 & 83253 &  3.6 & 83125 & 83095 & 83150 &  97 & 4\\
CGF~5-3  & Sb       & 3:39:34.31 &-35:27:22.0 & 18.48 & 47942 &  6.1 & 48050 & 47660 & 47996 &  57 & 1\\
CGF~5-4  & cE/GC?   & 3:39:35.92 &-35:28:24.9 & 19.05 &  1874 &  3.9 & 1863 &      &  1869 &  60 & 3\\
CGF~7-9  & (d?)Irr  & 3:39:51.83 & -35:33:05.4 & 19.72 &   &    & 1981 &         &  1981 &  72 & 4\\
CGF~7-10  & E/S0     & 3:39:52.51 & -35:34:46.9 & 19.81 &  &     & 73169 &        & 73169 &  49 & 4\\
CGF~7-8  & E?       & 3:39:54.43 & -35:38:51.7 & 19.72 & 22287 &  4.3 & 22060 &      & 22165 &  73 & 3\\
CGF~7-2  & E        & 3:40:00.81 & -35:35:27.7 & 18.62 & 55257 &  8.0 & 55360 &      & 55309 &  53 & 1\\
CGF~7-1  & E(pec)   & 3:40:04.39 & -35:36:52.6 & 17.16 & 36199 & 12.1 & 36193 &      & 36196 &  38 & 1\\
CGF~7-4  & Sb       & 3:40:16.15 & -35:33:14.2 & 19.03 & 36360 &  4.4 & 36404 &36202 & 36382 &  72 & 1\\
CGF~7-7  & E        & 3:40:20.04 & -35:38:27.6 & 19.44 & 96894 &  5.3 & 96945 &     & 96920 &  57 & 2\\
CGF~10-9  & E        & 3:42:00.93 &-35:25:43.7 & 19.42 & 46416 &  4.4 & 46653 &      & 46534 &  77 & 3\\
CGF~10-6  & Sc(BCD?) & 3:42:02.09 &-35:23:45.8 & 18.85 & 25725 &  3.3 & 25521 & 25612 & 25591 & 107 & 3\\
CGF~10-3  & Sa       & 3:42:03.83 &-35:26:47.3 & 18.41 & 54727 &  6.7 & 54714 &      & 54721 &  41 & 1\\
CGF~10-2  & E        & 3:42:05.42 &-35:23:58.1 & 17.90 & 57092 & 13.0 & 57095 &      & 57094 &  41 & 1\\
CGF~10-4  & S0       & 3:42:06.13 &-35:23:41.4 & 18.65 & 56663 &  4.8 & 56004 & 56632 & 56318 &  71 & 1\\
CGF~10-8  & Sb?      & 3:42:14.75 &-35:26:17.4 & 19.35 &       &      & 55053 & 55012 & 55033 & 123 & 1\\
FCC 274 & dE,N     & 3:42:17.28 &-35:32:20.8 & 16.50 &  1095 &  4.5 & 1058 &      &  1073 &  76 & 3\\
CGF~10-1  & Sb?      & 3:42:19.01 &-35:23:47.2 & 16.87 &       &      & & 16409 & 16409 &  40 & 2\\
CGF~10-15  & S        & 3:42:20.02 &-35:26:14.7 & 20.07 &       &      & 29577 &      & 29577 &  66 & 4\\
CGF~10-5  & Sc?      & 3:42:27.62 &-35:27:01.1 & 18.68 &       &      & & 38463 & 38463 & 120 & 4\\
\noalign{\smallskip}
\hline
\end{tabular}
\end{flushleft}
\end{table*}

\begin{figure}
\psfig{figure=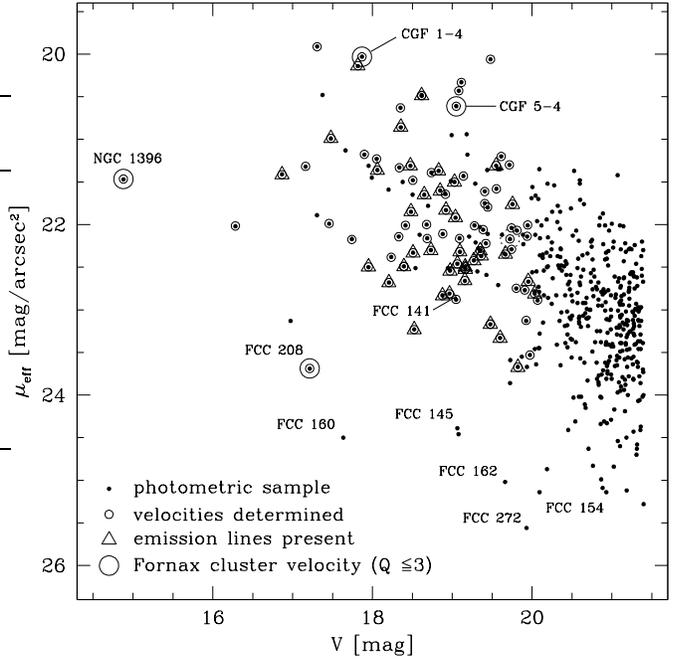,height=8.6cm,width=8.6cm
,bbllx=9mm,bblly=65mm,bburx=195mm,bbury=249mm}
\vspace{0.4cm}
\caption{In this magnitude -- surface brightness diagram all galaxies of
our photometric sample brighter than $V = 21.5$ mag are plotted. Galaxies
with velocity determinations are encircled with small circles. The background
galaxies that show emission lines are indicated with triangles. Only the
galaxies encircled with large circles are Fornax members.
The spectra of most dwarf galaxies
have a too low signal-to-noise for line identifications.
Note that the dE,Ns FCC 188, FCC 222, and FCC 274 are not
plotted, since they do not belong to our photometric sample
}
\end{figure}

94 radial velocities were measured (not including the velocities of the giant 
galaxies of Table 1).
The data are listed in Table 2 in order of increasing right ascension. The
first column is the catalog name from Paper~I.
Column 2 gives the galaxy type as determined by morphological appearance,
surface brightness profile (see Paper~I), and spectral type.
For the spectral classification the spectrophotometric
atlas of galaxies by Kennicutt (\cite{kenn}) was used for comparison.
Columns 3 -- 4 are the right ascension, declination (epoch 2000.0) and total V
magnitude from Paper~I.
Columns 6 -- 9 are the heliocentric velocity of the cross correlation technique
$v_{\rm cross}$, the R value, the velocity of the second determination method
$v_{\rm rvid}$, and the emission line velocity $v_{\rm em}$, if available.
In column 10 and 11
the final adopted velocity and a mean error of the different determinations
are given. In most of the cases, the final velocity is the mean of 
$v_{\rm cross}$ and $v_{\rm rvid}$.
In the cases where the emission line velocity is based on at
least 2 clearly identified lines it was also included.
Column 12 indicates the quality parameter Q of the velocity determination.
Table 2 is also available in electronic form at the CDS via anonymous
ftp to cdsar.u-strasbg.fr (130.79.128.5) or via 
http://cdsweb.u-strasbg.fr/Abstract.html.

We defined 4 classes of quality due to the clearness and evidence of the 
velocity
identification: (1) a clear and evident
identification, both methods agree very well, (2) very probable identification,
almost clear, (3) doubtful, but most probable value, and (4) very
doubtful, only a try.

\begin{table}
\caption{Statistics of the quality of velocity determinations}
\begin{flushleft}
\begin{tabular}{crrrr}
\hline\noalign{\smallskip}
Q & Fornax & cluster & others & total \\
\noalign{\smallskip}
\hline\noalign{\smallskip}
1 & 4 & 15 & 23 & 42 \\
2 & 0 & 2 & 10 & 12 \\
3 & 3 & 1 & 22 & 26 \\
4 & 1 & 1 & 12 & 14 \\
\hline
total & 8 & 19 & 67 & 94 \\
\noalign{\smallskip}
\hline
\end{tabular}
\end{flushleft}
\end{table}

Table 3 summarises the statistics of quality classes divided
into the following sub-samples: velocities
between 700 and 2500 km~s$^{-1}$ (Fornax cluster, not including the galaxies of 
Table 1),
velocities between 33000 and 34500 km~s$^{-1}$ (background cluster, see Sect.~5.3),
and all other background galaxies.

\begin{figure}
\vspace{0.3cm}
\psfig{figure=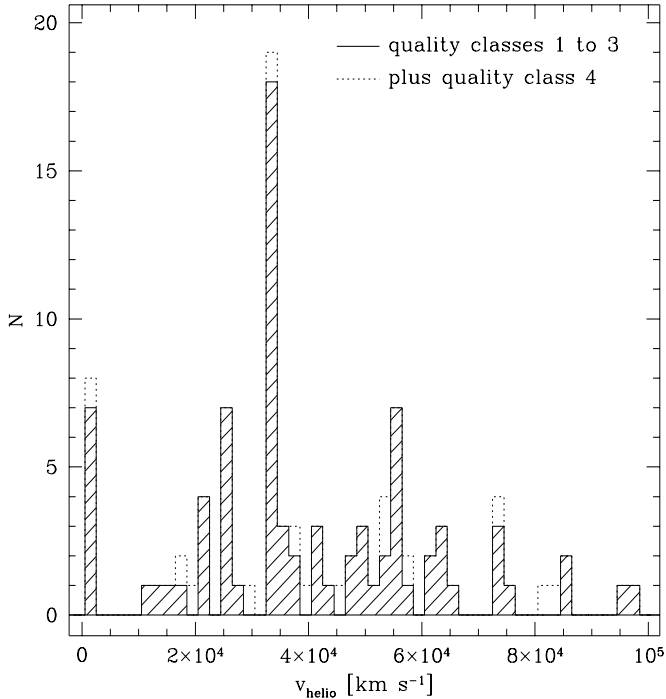,height=8.6cm,width=8.6cm
,bbllx=9mm,bblly=65mm,bburx=195mm,bbury=239mm}
\vspace{0.4cm}
\caption{Velocity histogram for all galaxies with velocity determinations.
The bin size is 2000 km~s$^{-1}$ starting at 500 km~s$^{-1}$.
The concentration of galaxies at about 33700 km~s$^{-1}$ belongs to a background
cluster just behind the center of the Fornax cluster
}
\end{figure}

In Figure 3 we show the selection of our sample in a 
$\mu_{\rm eff}$--$V_{\rm tot}$ diagram (see also Paper~I, Fig.~3 and 7). All 
velocity determinations of the quality classes
1 to 3 are indicated. The limits are $V_{\rm limit}
\simeq 20$ mag and $\mu_{\rm eff_{\rm lim}} \simeq 23.5$ mag~arcsec$^{-2}$.
The velocities of the fainter galaxies are mainly based on
emission line identifications. Due to the low surface brightness of most
dwarf galaxies (FCC numbers) the signal-to-noise of their spectra is too
low for line identifications. Among the galaxies with higher surface
brightness two objects with velocities consistent with Fornax members
were found.

The distribution of velocities in our sample is shown in Fig.~4. The first
bin represents the Fornax cluster velocities. The highest
velocity ($cz = 97000$ km~s$^{-1}$) corresponds to a redshift of $z = 0.3$.
A striking feature is the concentration of galaxies around 33700 km~s$^{-1}$ 
($z = 0.11$).
They belong to a background cluster behind the center of the 
Fornax cluster (see Sect.~5.3). Another group of galaxies with velocities around
26000 km~s$^{-1}$ is concentrated in a region about $12\arcmin$ east of 
NGC~1399 (field B1; see Paper~I). The galaxies around 55000 km~s$^{-1}$ are not
spatially correlated.

\subsection{Previously measured galaxies}

Our velocity sample includes 12 galaxies studied before.
Table 4 gives the different names, membership classifications, and previous
radial velocity measurements.

Irwin et al.~(\cite{irwi}) had no morphological classification for their 
targets,
but noted that their compilation of low surface brightness
galaxies, which they classified as members (Davies et al.~\cite{davi}), might be
contaminated by background galaxies in the magnitude range $17 < B < 19$ mag
and of central surface brightnesses of $22.0 < \mu_B < 22.5$ mag/arcsec$^2$.
Indeed, all galaxies in this range turn out to be background galaxies
(CGF~9-1, CGF~3-6,
and CGF~1-10). Concerning the FCC, all ``likely background'' galaxies
are proved to be true background galaxies (CGF~9-1, CGF~3-7, and CGF~10-2).
In addition, the likely Fornax member FCC 141 (CGF~9-5) has a velocity of
$v_{\rm helio} = 16845$ km~s$^{-1}$. Due to its blue color,
$(V - I) = 0.8$ mag, it was classified as irregular.
The definite FCC galaxies are confirmed to be members. Their velocities agree
well with previous results.

\begin{table*}
\caption{Cross references and comparison with velocities
to previous identifications}
\begin{minipage}{18.0cm}
\begin{flushleft}
\begin{tabular}{llcrrcrc}
\hline\noalign{\smallskip}
  & FCC\footnote{Fornax Cluster Catalog (Ferguson \cite{ferg89}), numbers 
  with ``B''
in front are from the background galaxy catalog} & m\footnote{membership
classes (Ferguson \cite{ferg89}): 1 = definite member, 2 = likely member, 
4 = likely background} & D\&I\footnote{Davies et al.~(\cite{davi})
and Irwin at al.~(\cite{irwi})} &
$v_{\rm lit}$ & Ref\footnote{sources of radial velocities: 1 = Jones \& Jones
(\cite{jone}), 2 = Held \& Mould (\cite{held}), 3 = Carter \& Malin 
(\cite{cart})} &
$v_{\rm helio}$\footnote{this paper} & Q \\
\noalign{\smallskip}
\hline\noalign{\smallskip}
CGF~9-5 & 141 & 2 &     & & & 16833$\pm$95 & 3\\
CGF~9-1 & B1016   & 4 & 231 & & & 50115$\pm$45 & 1\\
         & 188 & 1 &  76 & 999$\pm$38 & 2 & 945$\pm$67   & 3\\
CGF~4-1\footnote{CGF~4-1 = NGC~1396} & 202 & 1 &     & 894$\pm$120 & 1 & 882$\pm$39 & 1\\
CGF~3-7     & B1220   & 4 &     & & & 21835$\pm$29 & 1\\
CGF~3-2     & 208 & 1 & 257 & & & 1694$\pm$85 & 1\\
CGF~3-6     &         &   & 258 & & & 61365$\pm$53 & 1\\
CGF~2-2     &         &   &     & 33750$\pm$30 & 3 & 33869$\pm$58 & 1\\
CGF~1-10    &         &   & 266 & & & 64469$\pm$41 & 1\\
         & 222 & 1 &     & 953$\pm$150 & 1 & 852$\pm$52 & 1\\
CGF~10-2 & B1571   & 4 &     & & & 57094$\pm$41 & 1\\
         & 274 & 1 & 141 & & & 1073$\pm$76 & 3\\
\noalign{\smallskip}
\hline
\end{tabular}
\end{flushleft}
\end{minipage}
\end{table*}

Recently, Minniti et al.~(\cite{minn}) have measured radial velocities of 
globular clusters
around NGC~1399. Their brightest object, whose nature as GC was questionned, 
is the same as the nucleus-like
object CGF~1-4 (see next section). Their velocity of $v_{\rm helio} = 1459\pm52$ km~s$^{-1}$
agrees very well with our value: $v_{\rm helio} = 1485\pm38$ km~s$^{-1}$.

\subsection{New Fornax members}

\begin{table*}
\caption{Number counts of GCs in the bright end of the GCLF
from MC simulations. The total number of GCs is 5800. The bin centers are
given in absolute $V$ magnitudes.}
\begin{flushleft}
\begin{tabular}{ccllllll}
 & & & & & & \\
\hline\noalign{\smallskip}
function & $\sigma$  & -14.3 mag & -13.8 mag & -13.3 mag & -12.8 mag & -12.3
mag & -11.8 mag\\
\noalign{\smallskip}
\hline\noalign{\smallskip}
t5 & 0.9 & $0.6\pm0.8$ & $1.2\pm1.2$ & $1.7\pm1.0$ & $2.9\pm1.7$ & $4.7\pm2.2$
 & $8.8\pm3.1$ \\
t5 & 1.1 & $1.6\pm1.3$ & $2.6\pm1.5$ & $4.2\pm1.9$ & $6.2\pm2.7$ & $10.6\pm3.5$
 & $17.8\pm4.1$ \\
gauss & 1.2 & $0.0\pm0.0$ & $0.0\pm0.0$ & $0.0\pm0.0$ & $0.0\pm0.0$ &
$0.5\pm0.7$ & $2.6\pm1.7$ \\
gauss & 1.5 & $0.0\pm0.0$ & $0.1\pm0.3$ & $0.6\pm1.0$ & $2.3\pm1.7$ &
$5.8\pm2.7$ & $15.3\pm3.2$ \\
\noalign{\smallskip}
\hline
\end{tabular}
\end{flushleft}
\end{table*}

\begin{table*}
\caption{Line indices for Fornax members, given in magnitudes and
measured on the unfluxed spectra}
\begin{flushleft}
\small{
\begin{tabular}{lccccccc}
\hline\noalign{\smallskip}
Galaxy & Mg2 & MgH & Mgb & Fe5270 & Fe5335 & H$\beta$ & Gband\\
\noalign{\smallskip}
\hline\noalign{\smallskip}
NGC~1396 & 0.10$\pm$0.02 & 0.02$\pm$0.02 & 0.06$\pm$0.03 &
0.06$\pm$0.03 & 0.09$\pm$0.03 & 0.07$\pm$0.03 & 0.17$\pm$0.03 \\
FCC 188 & 0.13$\pm$0.09 & 0.00$\pm$0.07 & 0.18$\pm$0.16 &
... & 0.10$\pm$0.17 & 0.14$\pm$0.14 & 0.08$\pm$0.17 \\
FCC 222 & 0.06$\pm$0.07 & ... & 0.11$\pm$0.11 &
0.05$\pm$0.11 & 0.03$\pm$0.13 & 0.06$\pm$0.10 & 0.25$\pm$0.11 \\
CGF~1-4 & 0.26$\pm$0.02 & 0.07$\pm$0.02 & 0.17$\pm$0.04 & 0.10$\pm$0.03
 & 0.11$\pm$0.05 & 0.08$\pm$0.04 & 0.17$\pm$0.04 \\
CGF~5-4 & 0.08$\pm$0.06 & 0.01$\pm$0.05 & 0.06$\pm$0.11 & 0.10$\pm$0.08
 & 0.07$\pm$0.12 & 0.06$\pm$0.09 & ... \\
\noalign{\smallskip}
\hline
\end{tabular}
}
\end{flushleft}
\end{table*}

Three objects in our sample have velocities of the Fornax cluster, but
have not been identified as members yet.
Two of them, CGF~1-4 and CGF~5-4, are located at distances of $8\farcm3$ and 
$13\farcm6$ from 
NGC~1399, CGF~1-4 in the direction of NGC~1404 and CGF~5-4 in east direction. 
They are hardly resolved, have a circular shape, $(V - I)$ colors
of 1.1 and 1.0 mag, respectively, and a very high central surface brightness. 
These properties are typical for globular clusters as well as for nuclei of
nucleated dwarf ellipticals, or even for compact ellipticals. Adopting a
distance modulus to the Fornax cluster of $(m-M)_0 = 31.3$ mag, implying 
a distance of 18.2 Mpc,
(Kohle et al.~\cite{kohl}, recalibrated with new distances of Galactic GCs,
Gratton et al.~\cite{grat}, following Della Valle et al.~\cite{dell}), their 
absolute magnitudes are $M_V = -13.4$
mag and $M_V = -12.2$ mag, which is about 3 magnitudes fainter than M32
($M_V = -16.4$ mag) and 1 -- 2 magnitudes brighter than the
brightest GC of the investigated luminosity function (Kohle et al. \cite{kohl}).
For comparison, the brightest GCs in the central Virgo 
galaxy M87 have absolute V magnitude in the order of -11.5 mag (Elson \&
Santiago \cite{elso}; Whitmore et al.~\cite{whit}).

In order to investigate the possibility, whether this two objects might
be ``normal'' GCs in a very rich GCS, the number of GCs that populate
the bright end of the luminosity function (LF) of the GCS in NGC~1399 was 
estimated in Monte Carlo simulations.
As representation of the LF both a Gaussian and a t5-function
(see e.g. Kohle et al.~\cite{kohl}) with different dispersions $\sigma$ were 
adopted.In 100 runs 5800 GCs were randomly distributed. In Table 5
the number counts in 6 bright bins (bin width 0.5 mag) are given for
different functions and dispersions.
Very bright GCs with $M_V = -13.3$ mag can statistically
exist in a rich GCS, if the t5-function is representative for the bright
wing of the LF.

According to numerical simulations by Bassino et al. (\cite{bass})
nuclei of nucleated 
dwarf galaxies can survive the dissolution in the gravitational field of a
giant elliptical over the lifetime of the universe. They expect 
globular-cluster-like remnants and less concentrated remnants with
masses in the range between 2.8 to $7.4\cdot10^6M_{\odot}$ and tidal radii from
170 to 400 pc.

Assuming that the two nucleus-like objects have mass-to-light ratios
resembling GCs, their masses can be estimated.  
Adopting the relation of Mandushev et al.~(\cite{mand}), log$(M/M_\odot) = 
-0.431M_V 
+2.01$, we derive masses of $6.1\cdot10^7M_\odot$ and $1.9\cdot10^7M_\odot$
respectively.
This is one order of magnitude more massive than $\omega$ Centauri 
($2.5\cdot10^6M_\odot$), the most massive cluster in the Milky Way.
Alternatively, the mass of both objects can be compared with that of
M32. Nolthenius \& Ford (\cite{nolt}) derived a mass of about 
$8\cdot10^8M_\odot$
and a $M/L_B =$ 3-4 from velocity dispersion measurements. If we adopt
a mass-to-light ratio of 3.5 for the two compact objects, we get masses of
$6.3\cdot10^7M_\odot$ and $2.1\cdot10^7M_\odot$, more than one magnitude
less massive than M32.

The third object that has the velocity of the Fornax cluster, CGF~7-9, is
located about $6\arcmin$ north-east of NGC 1427A. Its velocity is
very uncertain (Q = 4). Nevertheless, in case of membership, this galaxy
belongs to the late-type dwarf population due to its blue color ($(V - I)
= 0.9$) and irregular shape.\\

\begin{figure}
\psfig{file=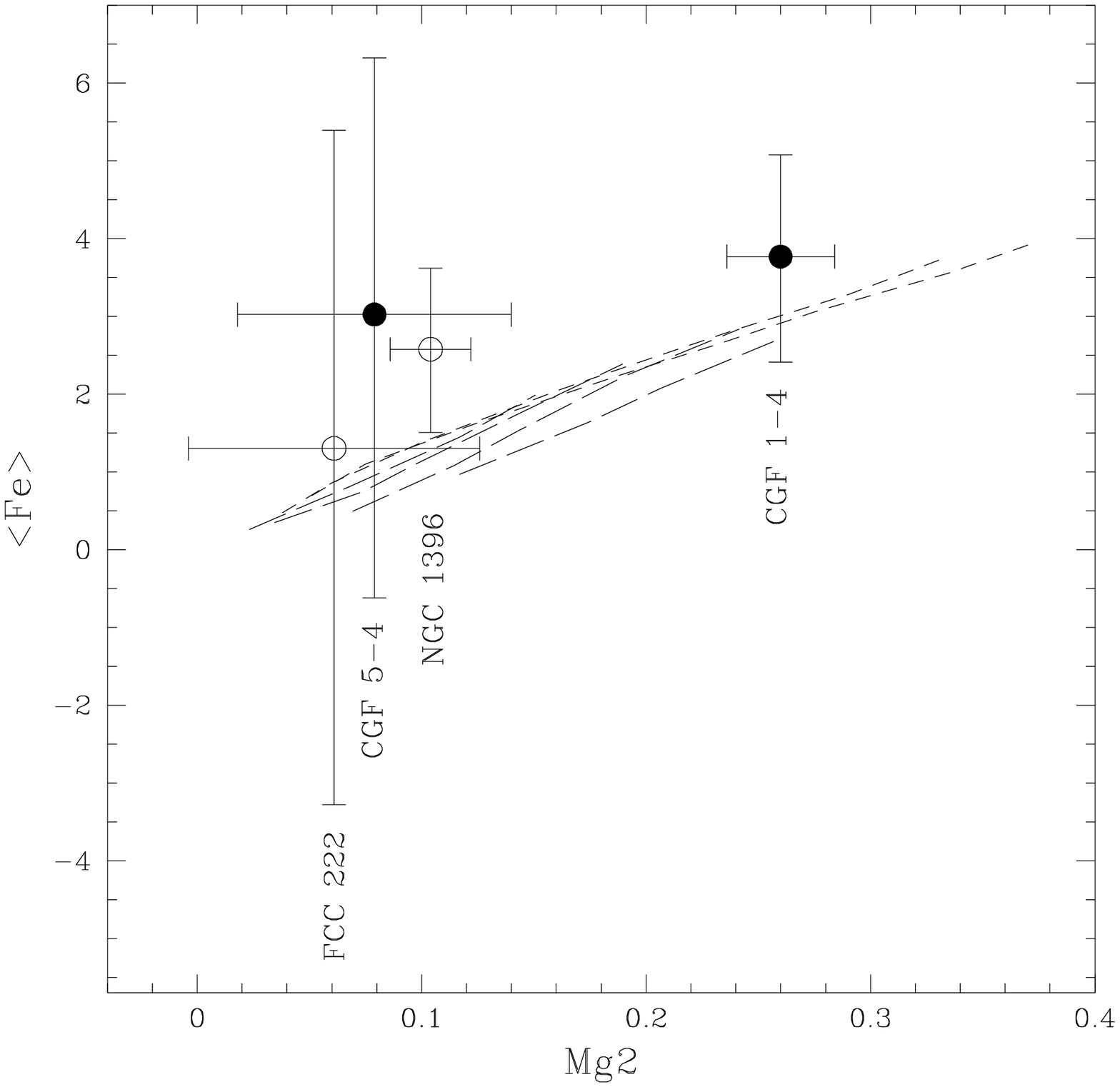,height=8.6cm,width=8.6cm
,bbllx=9mm,bblly=65mm,bburx=195mm,bbury=246mm}
\vspace{0.4cm}
\caption{Mg2 is plotted versus the equivalent width of
$<$Fe$>$ $=$ (Fe5270+Fe5335)/2. The filled circles are the nucleus-like
objects CGF~1-4 and CGF~5-4. Open circles are the dE,Ns FCC 122 and NGC 1396.
The dashed lines are tracks from population synthesis models for single
stellar populations (long dashed: Fritze-v.~Alvensleben \& Burkert \cite{frit},
short dashed: Worthey \cite{wort}). The age range is 1.5 to 17 Gyr,
the metallicity range $-2.0 < [{\rm Fe/H}] < 0.5$ dex.
}
\end{figure}

\subsubsection{Line indices of the Fornax members}

Metal abundances of the three dE,Ns NGC~1396, FCC 188, and FCC 222
and the two nucleus-like objects were measured in order to compare them with 
the properties of Milky Way, LMC and NGC~1399 GCs.
The procedure for the measurement of the line indices is described in
Brodie \& Huchra (\cite{brod}). The Lick/IDS bandpasses were used, as defined
by Burstein et al.~(\cite{burs}) and updated by Trager (\cite{trag}). In Table 6
the indices, given in magnitudes and measured on the unfluxed spectra, are 
summarised. In the Figs. 5, 6, and 7 the line indices of H$\beta$, $<$Fe$>$,
MgFe, and Mgb$\ast$Fe52 are converted into equivalent widths by the relation
$W_{\lambda}(I) = (\lambda_2 - \lambda_1)(1 - 10^{-I/2.5})$, where $\lambda_2$
and $\lambda_1$ are the maximum and minimum wavelengths of the bandpass.
The errors were estimated from the photon noise in the bandpasses,
$\sigma_{\rm P} = (\sigma^2_{\rm C1} + \sigma^2_{\rm L} +
\sigma^2_{\rm C2})^{1/2}$, where $\sigma_{\rm L}$, $\sigma_{\rm C1}$, and
$\sigma_{\rm C2}$ are the statistical errors in the line and adjacent
continuum bandpasses (see Brodie \& Huchra \cite{brod}). The statistical 
error in the
flux in a bandpass is $\sigma = (O + S)^{1/2}/O$, where O and S are the total
accumulated counts in the bandpass in the object and sky, respectively.

\begin{figure}
\psfig{file=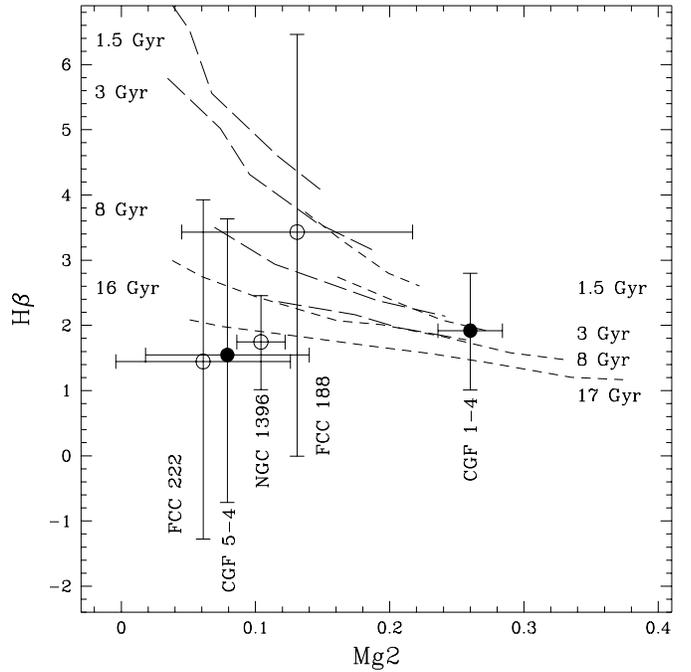,height=8.6cm,width=8.6cm
,bbllx=9mm,bblly=65mm,bburx=195mm,bbury=246mm}
\vspace{0.4cm}
\caption{H$\beta$ versus Mg2 with tracks from Fritze-v.~Alvensleben \&
Burkert (\cite{burk}) (long dashed lines, ages on the left,
metallicity between $Z = 0.001$ and $Z = 0.04$) and from Worthey (\cite{wort})
(short dashed lines, ages on the right, metallicity between $[{\rm Fe/H}] = -2.0$
to 0.5 dex)
}
\end{figure}

In Fig.~5 and Fig.~6 we plotted Mg2 versus the equivalent
widths of $<$Fe$>$ ($=$ (Fe5270 + Fe5335)/2) and H$\beta$,
respectively. The dE,N FCC 188 was omitted in the Mg2-$<$Fe$>$ plot, since the
iron lines in this spectra are too weak. Also plotted are the relations for
the indices as derived from population synthesis models for single stellar
populations. The long dashed lines are the models from Fritze-v.~Alvensleben
\& Burkert (\cite{frit}) for 1.5 to 16 Gyr old populations, with metallicities
between $Z = 0.001$ and $Z = 0.04$. Short dashed lines are Worthey's 
(\cite{wort})
models for 1.5 to 17 Gyr and $[{\rm Fe/H}] = -2.0$ to 0.5 dex.

The larger errors (due to the low signal-to-noise of the spectra) do not allow
an age separation of the objects. Notice that the nuclei of the dE,Ns
and CGF~5-4 fall in the range of metal-poor old single stellar populations.
Similarly, metal-poor GCs in the Milky Way and M31 (Burstein et al.~\cite{burs},
Brodie \& Huchra \cite{brod}) as well as in NGC~1399 (Kissler-Patig et 
al.~\cite{kiss}) are
located in this region of the plot. On the other hand, in this diagram CGF~1-4 
is clearly separated from these objects
and shows a more metal-rich stellar population. Its line indices are comparable
to those of the metal-rich GCs in the MW, M31, and NGC~1399
as well as the line indices of the center of M32 (Burstein et al.~\cite{burs}),
which shows a slight enhancement of H$\beta$ compared to GCs.

\begin{figure}
\psfig{file=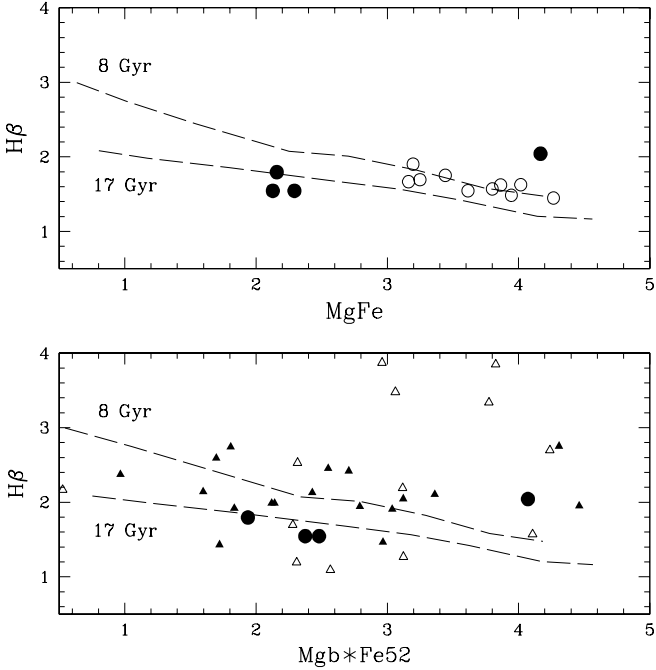,height=8.6cm,width=8.6cm
,bbllx=9mm,bblly=65mm,bburx=195mm,bbury=246mm}
\vspace{0.4cm}
\caption{[MgFe] versus H$\beta$ for the dwarf galaxies (filled circles) and
Fornax ellipticals (open circles) (upper panel), and [Mgb$\ast$Fe52] (see
text) versus H$\beta$ for Virgo dwarf galaxies (open triangles) and NGC~1399
globular clusters (solid triangles) (lower panel). For reference, we
show two models from Worthey (\cite{wort}) for 17 and 8 Gyr,
spanning a metallicity range [Fe/H] from $-2.0$ to 0.5 dex (dashed
lines)}
\end{figure}

Adopting the Mg2--$[{\rm Fe/H}]$ relation from Brodie \& Huchra
(\cite{brod}), the 4 more metal-poor objects have metallicities
that range between $[{\rm Fe/H}] \simeq -1.6$ and $-0.9$ dex.
This agrees well with the spectroscopic measurements of 10 dE,Ns in Fornax
by Held \& Mould (\cite{held}).
The metallicity of CGF~1-4 is $[{\rm Fe/H}] \simeq 0.4$ dex, when adopting the
Mg2-[Fe/H]
relation by Brodie \& Huchra (\cite{brod}), or $[{\rm Fe/H}] \simeq -0.1$ 
dex considering
the non-linear behaviour of Mg at higher metallicities (Worthey \cite{wort},
Kissler-Patig et al.~\cite{kiss}).

Finally, we show in Fig.~7 a comparison between the abundances of the dwarf
galaxies and elliptical galaxies in Fornax measured by Kuntschner \& Davies
(\cite{kunt}) (upper panel), as well as dwarf galaxies in Virgo
(taken from Huchra et al.~\cite{huch}) and globular clusters in NGC~1399
(taken from Kissler-Patig et al.~\cite{kiss}) (lower panel). The
[MgFe] index ($\sqrt{{\rm Mgb} \times <{\rm Fe}>}$) was plotted versus 
H$\beta$, following Kuntschner \& Davies. The Virgo dwarfs have no Fe5335 index 
measured. For them and the NGC~1399 GCs $\sqrt{{\rm Mgb} \times {\rm Fe5270}}$
(labeled [Mgb$\ast$Fe52]) was plotted versus H$\beta$.
In both panels models from Worthey (\cite{wort}) for 17 and 8 Gyr, spanning a 
metallicity range [Fe/H] from -2.0 to 0.5 dex, are shown for reference.
The large errors in our measurement prevent a detailed comparison;
age differences cannot be discriminated within a factor of 2 or 3.
However, we note that, as expected,
the dwarf galaxies (except CGF~1-4) are less metal rich than the
giant ellipticals and consistent with having similar ages.
The comparison with the Virgo dwarf galaxies and the NGC~1399 GCs shows 
that all the Fornax dwarfs (including CGF~1-4) fall in ranges span by them.
However, CGF~1-4 is more metal
rich than the bulge-like GCs in NGC~1399, and could belong
to the ``very metal rich'' group of GCs found by Kissler-Patig
et al.~(\cite{kiss}). In this case, it might also be somewhat younger, which
would increase its mass-to-light ratio and reduce the estimated mass.
But CGF~1-4 and CGF~5-4 remain puzzling objects; it can not yet be decided
whether they are nuclei of disrupted dwarf galaxies, cEs, or true extremely
massive GCs. Further spectroscopic observations with a higher 
signal-to-noise are needed to uncover the nature of these objects.

\begin{figure}[t]
\psfig{figure=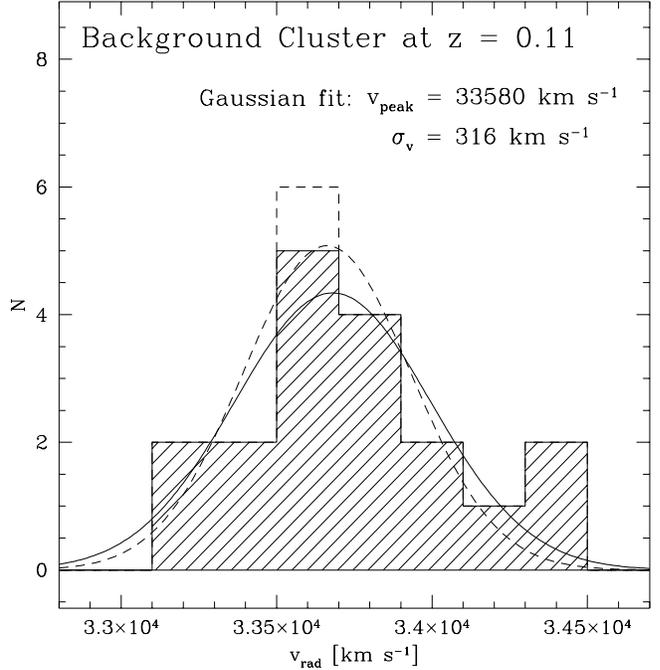,height=8.6cm,width=8.6cm
,bbllx=9mm,bblly=65mm,bburx=195mm,bbury=246mm}
\vspace{0.4cm}
\caption{Velocity histogram for background cluster galaxies. The velocity
dispersion is typical for a relatively poor galaxy cluster
}
\end{figure}
 
\subsection{Background cluster at $z = 0.11$}

19 galaxies with velocities around 33700 km~s$^{-1}$ (Fig.~8) were found. 
The velocity dispersion is about 360 km~s$^{-1}$ which is typical for poor
clusters (e.g. den Hartog
\& Katgert \cite{denh}). The ratio of early type (E+S0) to late type (S+Irr) 
giant 
galaxies is about 1.1. It is slightly lower than in the Fornax cluster.
The spatial distribution of the 19 galaxies is shown in Fig.~9 (bold hexagons).
It matches well the density distribution of the fainter galaxies down to
$V = 21.5$ mag; see Fig.~9 in Paper~I. 
Since no countable amount of new Fornax dwarf galaxies was found, we 
conclude that the excess population of galaxies near NGC~1399 mainly
belongs to the background galaxy cluster. 

Assuming a Hubble constant of $H_0 = 75$ km$\cdot$s$^{-1}\cdot$Mpc$^{-1}$
the distance to the cluster is 450 Mpc, or $(m - M)_V = 38.3$ mag.
The brightest cluster galaxy (CGF~1-1), located $1\farcm1$ south of NGC~1399, 
would then have an absolute luminosity comparable to NGC~1399 ($M_V = -22.1$
mag). Note that the K-corrections at this redshift are in the order of 0.15 mag 
in $V$ for old stellar populations and 0.05 mag for late-type spirals
(Coleman et al.~\cite{cole}). 
The radial surface brightness profile of this galaxy (see Paper~I, Fig.~4)
follows a de 
Vaucouleur profile in the inner part and becomes significantly flatter outside
a radius of $7\farcs0$ ($\simeq 16$ kpc), which is typical for a cD galaxy 
(e.g. Schombert \cite{scho}). 

\begin{figure}[t]
\psfig{figure=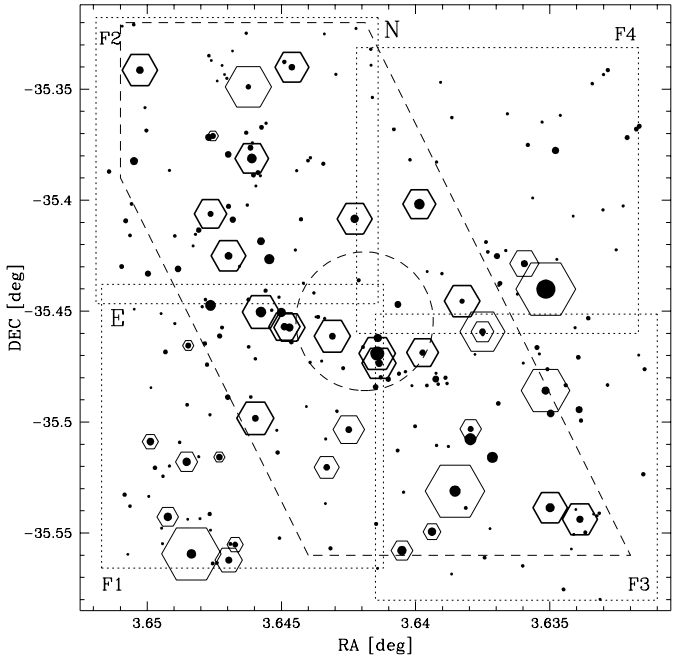,height=8.6cm,width=8.6cm
,bbllx=9mm,bblly=65mm,bburx=195mm,bbury=249mm}
\vspace{0.4cm}
\caption{Position of the background cluster galaxies in the central Fornax
CCD fields. The dashed circle is the central galaxy NGC~1399. The points are
galaxies brighter than $V = 21$ mag, the larger
the point the brighter the galaxy. The hexagons indicate the velocities of the
galaxies, the bold ones being velocities around 33700 km~s$^{-1}$. The larger
the hexagons around a galaxy the smaller is its velocity.
The dashed trapezium-like region was used for the analysis of the luminosity
and color distribution (see text)
}
\end{figure}

The luminosity and color distribution of the galaxies were analyzed inside a
trapezium-like region that encloses the member galaxies of the background
cluster (see Fig.~9). 
The distribution of an arbitrary sample of field galaxies around the 
other Fornax giant galaxies was subtracted from the distribution of the cluster
galaxies.

A dip in the luminosity distribution at $V = 19.5$ was found that
corresponds to an absolute magnitude of about $M_V = -18.8$ mag at a redshift
of $z=0.11$. This is the luminosity range where the Gaussian shaped giant
galaxy luminosity function decreases and the counts of dwarf galaxies start 
to rise (see e.g. Jerjen \& Tammann \cite{jerj}). Of course, the dwarf galaxy counts 
are severely uncomplete at the cluster distance
due to their very small angular diameters below our resolution limit
($1\farcs5 \simeq 3.3$kpc). However, at the fainter counts, an excess 
population of blue, $(V - I) < 0.8$ mag, dwarf galaxies was found, which
probably represents a population of star-forming dwarf irregulars in the
cluster. Most of these galaxies are clustered around the brightest cluster 
galaxy (CGF~1-1) and around a bright elliptical in the north-east of the 
cluster (CGF~2-2).

The galaxy CGF~2-2 is already known as a narrow-line radio galaxy (Carter \& 
Malin \cite{cart}).
It is correlated with the radio source PKS 0336$-$35.
Its spectrum shows strong O\,{\sc ii}, Ne\,{\sc iii}, and Ne\,{\sc v} emission
lines that  Carter \&
Malin explained by a very bright high-excitation emission-line nucleus.
Some of the blue faint galaxies in the direct environment of this galaxy 
have knots and tails.
We propose that CGF~2-2 is the center of an interacting subgroup of galaxies
within the background cluster.

In further investigations, this cluster might be used to constrain the amount 
of extincintion within the central part of the Fornax cluster.

\section{Summary}

Radial velocities of 94 objects in central fields of the Fornax cluster
have been measured with low resolution spectroscopy.
Our spectroscopic sample is limited in absolute magnitude at about
$V = 20.0$ mag and in peak surface brightness at about $\mu_{\rm peak} = 22.5$ 
mag~arcsec$^{-2}$. Most of the velocity determinations of the fainter galaxies
are based on strong emission lines of late type galaxies.

Eight objects were identified as Fornax members due to their radial velocities.
Five of them are listed as nucleated dwarf ellipticals in the FCC.
With one exception the membership and/or background classification by
Ferguson (\cite{ferg89}), which is based on galaxy morphology, agrees very 
well with our results.
Among the three ``new'' Fornax members, there are two that have photometric
properties that can be explained by a very bright GC
as well as by a compact elliptical like M32.
Another explanation might be that these objects represent the nuclei of
dissolved dE,Ns. The measurement of line indices has shown that the brighter
one of these objects has a solar metallicity, whereas the other nucleus-like
object as well as the nuclei of the measured dE,Ns have line indices that
are similar to those of old metal-poor GCs.
It would be interesting to investigate, whether there are more objects of this
kind hidden among the high surface brightness objects in the central
Fornax cluster.

Among the background galaxies we found a concentration of galaxies with
velocities around $v_{\rm helio} = 33700$ km~s$^{-1}$. These 19 galaxies are
spatially concentrated near the central Fornax galaxy NGC~1399 as well.
Their velocity dispersion, $\sigma_v = 316$ km~s$^{-1}$, is typical for a 
relatively
poor galaxy cluster. The brightest member galaxy, located $1\farcm1$ south of 
NGC~1399, has a similar absolute magnitude and extended cD
profile as NGC~1399.
We estimated that all excess galaxies in the central Fornax fields, except
the known and ``new'' Fornax members, most probably belong to the
background cluster just behind the Fornax center.
%
%%%%%%%%%%%%%%%%%%%%%%%%%%%%%%%%%%%%%%%%%%%%%%%%%%%%%%%%%%%%%%%%%%%%%%%%%%%
%%%%%%%%%%%%%%%%%%%%%%%%%%%%%%%%%%%%%%%%%%%%%%%%%%%%%%%%%%%%%%%%%%%%%%%%%%%
%%%%%%%%%%%%%%%%%%%%%%%%%%%%%%%%%%%%%%%%%%%%%%%%%%%%%%%%%%%%%%%%%%%%%%%%%%%

\acknowledgements
We thank Harald Kuntschner for providing us an electronic version of his
data on elliptical galaxies in Fornax. We also thank the referee H.C.~Ferguson
for his useful comments which improved the paper.
This research was partly supported by the DFG through the Graduiertenkolleg
`The Magellanic System and other dwarf galaxies' and through
grant Ri 418/5-1 and Ri 418/5-2. LI thanks Fondecyt Chile for support
through `Proyeto FONDECyT 8970009' and for a 1995 Presidential Chair in
Science.

\enddocument